\documentclass{emulateapj}
\usepackage{subfigure,lscape}

\newcommand{\AAi}{\mathrm{\AA}} %Improved angstroms
\newcommand{\dz}{\Delta z}

\newcommand{\dzrange}{\Delta z < 0.1, 0.2, {\mathrm{and~}} 0.3}
\newcommand{\dzdef}{|z_{\mathrm{spec}} - z_{\mathrm{phot}}|}

\newcommand{\zphot}{z_{\mathrm{phot}}}
\newcommand{\zspec}{z_{\mathrm{spec}}}

\slugcomment{Accepted to ApJ}
\shorttitle{Probabilistic Photometric Redshifts}
\shortauthors{Ball et al.}

\begin{document}

\title{Robust Machine Learning Applied to Astronomical Datasets III: Probabilistic Photometric Redshifts for Galaxies and Quasars in the SDSS and GALEX}
\author{Nicholas M. Ball\altaffilmark{1}, Robert J. Brunner\altaffilmark{1,2}, Adam D. Myers\altaffilmark{1}, \\ Natalie E. Strand\altaffilmark{3}, Stacey L. Alberts\altaffilmark{1}, David Tcheng\altaffilmark{2}}
\altaffiltext{1}{Department of Astronomy, MC-221, University of Illinois, 1002 W. Green Street, Urbana, IL 61801, USA}
\altaffiltext{2}{National Center for Supercomputing Applications, MC-257, 1205 W. Clark St, Urbana, IL 61801, USA}
\altaffiltext{3}{Department of Physics, MC-704, University of Illinois, 1110 W. Green Street, Urbana, IL 61801, USA}
\email{nball@astro.uiuc.edu}

\begin{abstract} %297 words
We apply machine learning in the form of a nearest neighbor instance-based algorithm (NN) to generate full photometric redshift probability density functions (PDFs) for objects in the Fifth Data Release of the Sloan Digital Sky Survey (SDSS DR5). We use a conceptually simple but novel application of NN to generate the PDFs---perturbing the object colors by their measurement error---and using the resulting instances of nearest neighbor distributions to generate numerous individual redshifts. When the redshifts are compared to existing SDSS spectroscopic data, we find that the mean value of each PDF has a dispersion between the photometric and spectroscopic redshift consistent with other machine learning techniques, being $\sigma = 0.0207 \pm 0.0001$ for main sample galaxies to $r < 17.77$ mag, $\sigma = 0.0243 \pm 0.0002$ for luminous red galaxies to $r \lesssim 19.2$ mag, and $\sigma = 0.343 \pm 0.005$ for quasars to $i < 20.3$ mag. The PDFs allow the selection of subsets with improved statistics. For quasars, the improvement is dramatic: for those with a single peak in their probability distribution, the dispersion is reduced from 0.343 to $\sigma = 0.117 \pm 0.010$, and the photometric redshift is within 0.3 of the spectroscopic redshift for $99.3 \pm 0.1$\% of the objects. Thus, for this optical quasar sample, we can virtually eliminate `catastrophic' photometric redshift estimates. In addition to the SDSS sample, we incorporate ultraviolet photometry from the Third Data Release of the Galaxy Evolution Explorer All-Sky Imaging Survey (GALEX AIS GR3) to create PDFs for objects seen in both surveys. For quasars, the increased coverage of the observed frame UV of the SED results in significant improvement over the full SDSS sample, with $\sigma = 0.234 \pm 0.010$. We demonstrate that this improvement is genuine and not an artifact of the SDSS-GALEX matching process.
\end{abstract}


\section{Introduction} \label{Sec: Intro}

Advances in CCD and other technologies are enabling modern wide-field surveys to provide high quality photometry for ever-increasing numbers of astronomical objects \citep[e.g.,][]{kron:surveys,reshetnikov:surveys,lawrence:surveys}. Comparable advances in multifiber spectrographs are enabling similarly increasing numbers of spectra to be taken \citep[e.g.,][]{lahav:zsurveys,yip:specsurveys}. However, due to the increased integration time required to obtain a meaningful spectrum to a given depth compared to an image, the number of spectra available typically lags the number of images by more than an order of magnitude. Given the importance of the physical information contained within a spectrum compared to that within an image, for example, much more accurate diagnostics of an object's type and its redshift, any comparable information that can be obtained from the image is of great importance. With the much larger numbers of objects for which photometry is available, for applications that do not require high resolution spectra this information can even surpass the spectra in terms of statistical significance \citep[e.g.,][]{blake:photozcosmology}.

However, for many applications, it is vitally important to know not only the photometric information, but also its relative accuracy within the dataset for each object. Typically, this might be achieved by, for example, providing an error on a measured magnitude, or an estimated Gaussian dispersion on a photometric redshift. In general, however, one would like to utilize the full {\it probability density function} (PDF) within analyses, so that one can exclude objects which do not meet specific criteria, or fold the information into the analysis.

An area in which PDFs are of particular utility is photometric redshifts. For many purposes, provided that they are reasonably accurate, the final raw accuracy of a redshift estimate is not vitally important, provided that the error distribution is well known. Photometric redshifts, particularly those of quasars, are known to suffer from a percentage of `catastrophic' failures, e.g., \citet{budavari:qsophotoz,richards:qsophotoz,weinstein:sdssqsophotoz,wu:qsophotoz}, in which the derived value is very different from the true value, e.g., $z \sim 0.7$ instead of $z \sim 2.2$. PDFs can help to minimize these because in many cases the PDF for such an object will contain two or more peaks in the redshift probability function.

Previous work on PDFs for photometric redshifts has concentrated on their derivation using a color-redshift relation, derived either empirically or from spectroscopic templates. Examples in which PDFs or $\chi^2$ distributions for objects are shown include
\citet{lanzetta:photoz,fernandezsoto:photoz,kodama:photoz,benitez:photoz,bolzonella:photoz,firth:annphotoz}, and \citet{brodwin:iracphotoz} for galaxies, and \citet{budavari:qsophotoz,richards:qsophotoz,weinstein:sdssqsophotoz}, and \citet{wu:qsophotoz} for quasars.

In this paper, we utilize objects with spectra from the Fifth Data Release \citep[DR5;][]{adelmanmccarthy:dr5} of the Sloan Digital Sky Survey \citep[SDSS;][]{york:sdss} to train a nearest-neighbor instance-based machine learning algorithm, and perform blind tests to assess the utility of the method in assigning PDFs. We present results for main sample galaxies \citep[MSGs;][]{strauss:mainsample}, luminous red galaxies \citep[LRGs;][]{eisenstein:lrgsample}, and quasars \citep{richards:qsosample}. Each of these samples have successively lower sample densities, but probe larger cosmic volumes. With our approach, it is also possible to generate PDFs for the entire SDSS photometric database. Similar work was carried out for classification probabilities by \citet{ball:dtclassification} for the quantity P(star,galaxy,neither-star-nor-galaxy). However, such an effort is beyond the scope of the current paper, as we work solely with objects for which spectra are available.

In addition to the SDSS, we cross-match the SDSS data to the Third Data Release \citep[GR3;][]{morrissey:galexgr3} of the Galaxy Evolution Explorer All-Sky Imaging Survey \citep[GALEX AIS;][]{martin:galex}. This provides an additional two bands in the near- and far-UV, giving useful information by extending the SED coverage, e.g., so that at $z < 0.3$ we can potentially sample information for quasars from both the \ion{Mg}{2} line in the UV and H$\alpha$ in the optical.

\section{Data} \label{Sec: Data}

We utilize data from the SDSS DR5 and the GALEX AIS GR3. In the SDSS, we select primary non-repeat observations of objects with spectra classified as galaxies and quasars ({\tt specClass} = {\tt galaxy}, {\tt qso} or {\tt hiz\_qso}) in the {\tt specObj} view of the Catalog Archive Server (CAS). In GALEX we select photometric objects with {\tt primary\_flag} = 1 from its similar CAS interface. All object attributes and errors used are from these sources. The data are retrieved via SQL queries.

In the SDSS, the object attributes retrieved are the magnitudes, $ugriz$, the associated errors derived from photon statistics \citep{stoughton:edr}, and the spectral type. The SDSS imaging covers the wavelength range $3000\AAi - 10,500\AAi$, and the spectra $3800\AAi - 9200\AAi$. Each magnitude is measured in four different apertures: {\it PSF}, {\it fiber}, {\it Petrosian}, and {\it model}; and we require all magnitudes to be within the range $0 < {\mathrm{\it mag}} < 40$, and magnitude errors to be within $0 < {\mathrm {\it magErr}} < 10$. Much tighter cuts could reasonably be applied; but we simply wish to eliminate extreme outlying values that are entirely unphysical (e.g., -9999) as they can cause instability in the learning algorithm. We also note that less outlying values should be easily accounted for by the learning process. Throughout this work, the SDSS magnitudes are corrected for Galactic extinction using the dust maps of \citet{schlegel:dustmaps}, and the GALEX magnitudes are corrected using the $B-V$ ({\tt e\_bv}) term inferred from these maps using the standard formula of \citet{cardelli:extinction}.

We subdivide the objects into three samples: main sample galaxies, luminous red galaxies, and quasars. Each sample is subject to additional cuts as appropriate. For MSGs (${\tt specClass} = 2$), we require, following \citet{strauss:mainsample},  Petrosian magnitude $r < 17.77$, ${\tt zWarning} = 0$, ${\tt zStatus} > 2$, and ${\tt zConf} > 0.85$. For LRGs, we apply the selection criteria of \citet{eisenstein:lrgsample}, resulting in ${\tt specClass} = 2$, ${\tt primTarget} = {\tt TARGET\_GALAXY\_RED}$, $z > 0.2$, ${\tt zWarning} = 0$, ${\tt zStatus} > 2$, and ${\tt zConf} > 0.85$. For quasars, we require ${\tt specClass} = 3$ or 4, ${\tt zWarning} = 0$, and ${\tt zStatus} > 2$. We remind the reader that extinction-corrected magnitudes are used throughout. The resulting numbers of objects are 413,361 MSGs, 66,268 LRGs, and 55,743 quasars. The quasar sample is the same as that of \citet[][hereafter B07]{ball:ibphotoz}, with the loss of three objects due to our additional restriction on the magnitude error.

\begin{figure*}
\figurenum{1}
\plotone{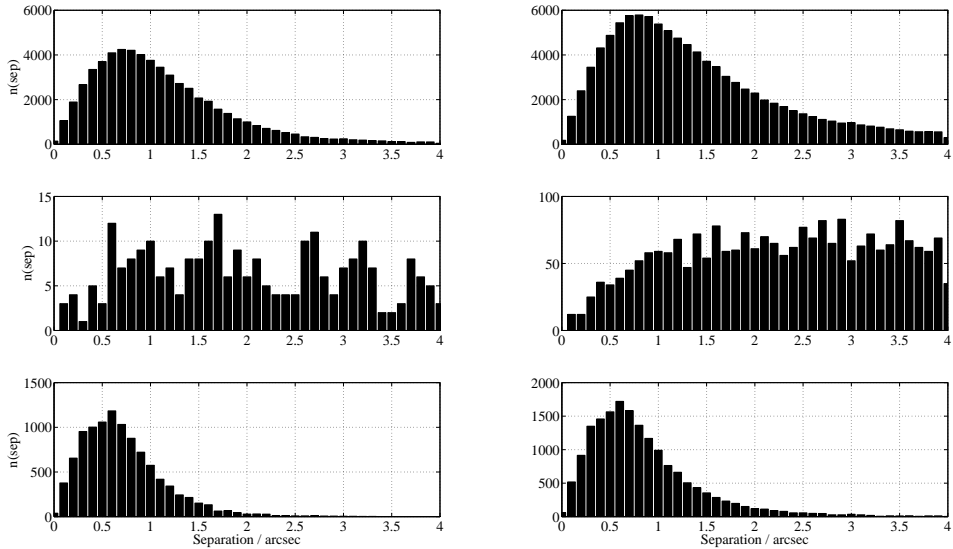}
\caption{Histograms of separations in SDSS-GALEX cross-matches. The rows show main sample galaxies, luminous red galaxies, and quasars respectively. The left-hand column shows FUV+NUV, the right-hand column shows NUV-only. \label{Fig: Separation}}
\end{figure*}

The SDSS samples are cross-matched to the primary photometric objects in the {\tt photoObjAll} view of the GALEX AIS GR3 database, using an RA+decl. tolerance (a distance on the sky) of 4 arcsec. This adds the near-UV ($1750\AAi - 2750\AAi$) and far-UV ($1350\AAi - 1750\AAi$) bands. We require the match to be unambiguous, in the sense that no SDSS object is within 4 arcsec of more than one GALEX object. Figure \ref{Fig: Separation} shows histograms of the object separations. The majority of these are much smaller than 4 arcsec, indicating that our tolerance is reasonable. For the GALEX photometry, we construct samples requiring a detection in both the near and far-UV bands (${\tt band} = 3$), and a second set of samples just requiring detection in the near-UV band (${\tt band} = 1$). The latter are constructed because the samples are considerably larger, but still incorporate some UV information. We again require magnitudes in the range 0--40, and the flags {\tt fuv\_artifact} and {\tt nuv\_artifact} are required to be 0. The resulting matches consist of 59,845, 256, and 10,328 objects for MSGs, LRGs, and quasars respectively in near- and far-UV; and 100,826, 2316, and 17,110 objects for near-UV only. Several qualitative features of these matches are as expected: (1) the near-UV-only samples are larger because they only require one detection not two; (2) there are very few matches to LRGs in GALEX, because LRGs are dominated by massive red early-type galaxies with little ongoing star formation; (3) the size of the quasar matched far- and near-UV sample increases in the same proportion to the size of the GALEX dataset as a whole compared to that obtained for GALEX GR2 by B07; and (4) there are few quasar matches beyond $z \sim 2$, because the Lyman limit at $912\AAi$ has redshifted out of the GALEX bands.

Following B07, in addition to the SDSS and SDSS+GALEX datasets, we also analyze the SDSS+GALEX sample of objects, but using only SDSS features. As in B07, these datasets are referred to as {\it GALEX-SDSS-only}, and they enable us to quantify the level of improvement in SDSS+GALEX seen from the addition of the GALEX UV features, as opposed to possible improvement due to the sample only containing luminous quasars that appear in both SDSS and GALEX.

%ADM: perh. discuss dropouts: objects in SDSS and should be in GALEX but are too faint. NMB: Have mentioned the z cutoff, but don't know about too-faint-in-GALEX.

\section{Algorithms} \label{Sec: Algorithms}

We apply the NN instance-based learning to each of the datasets. Full details of the NN method and its extension to $k$-nearest neighbors ($k$NN) are given in B07, or in, e.g., \citet{aha:ib,witten:datamining}, or \citet{hastie:learning}. Briefly, the method requires a set of training features for each object and a target property. The algorithm then compares the position in feature space of each new object in the testing set to the training set, and assigns the target property of the nearest training set object. This may be generalized to a weighted sum of nearest neighbors, i.e., $k>1$. The method is computationally intensive\footnote{For example, to generate the MSG PDFs requires 172048~s on 100 nodes for $n_{\mathrm{PDF}}~n_{\mathrm{galaxy}}~n_{\mathrm{validation}} = 100 \times 82672 \times 10$ galaxies, giving 4.8 galaxies $s^{-1}$.}, but we are able to exploit its full power by utilizing nationally allocated, peer-reviewed time on the Xeon Linux cluster Tungsten at the National Center for Supercomputing Applications (NCSA), and the Java environment Data-to-Knowledge \citep{welge:d2k}.

Throughout this paper, for the SDSS data the training features are the 4 colors $u-g$, $g-r$, $r-i$, and $i-z$ in the four magnitude types {\it PSF}, {\it fiber}, {\it Petrosian}, and {\it model}. This results in 16 training features. In B07, genetic algorithms were used to investigate subsets of these parameters in a systematic way; however, no subset was found that resulted in significant improvement, and indeed many subsets were worse. Preprocessing the training features with principal component analysis may remove some redundancy and save computational time, but given the aforementioned B07 result, we elected to simply use the full 16 colors throughout, in the spirit of using the full information available.

The target property is the spectroscopic redshift, which we regard as being correct as any error on this value is expected to be small compared to the photometric redshift error. When cross-matched to the GALEX data, the addition of the far-UV (FUV) and near-UV (NUV) bands gives the additional colors FUV-NUV, and four instances of NUV-$u$, resulting in 21 training features. The GALEX-SDSS-only sets contain the same objects as the SDSS+GALEX, but with just the assigned 16 SDSS features used.

As in B07, we standardize the training features such that each has a mean of 0 and a variance of 1. We test the performance of the algorithm by splitting each dataset into a training set, consisting of an 80\% random subsample of the data, and a blind test set, consisting of the remaining 20\%. The blind test set does not overlap with the training set, and this represents a realistic measure of the performance of the algorithm on unseen data within the same color regime. We perform 10-fold repeated holdout validation on each dataset, i.e., 10 different training:blind test splits, and quote the mean and standard deviation of the results for each.

The NN method as described, produces a single scalar-valued photometric redshift for each object (for each training:blind split). Therefore, to generate a probability density function in redshift for each object, we {\it perturb} the values of the training and blind features according to the given error on each feature. We assume the errors are Gaussian, and the perturbation is applied to the magnitude before the color is derived. This assumes that there is negligible covariance between the magnitudes. \citet{scranton:sdsscovariance} show that in fact this is not necessarily the case with the available SDSS magnitude errors. However, these errors are underestimated, which cancels the covariance at the 10-20\% level. We therefore use the values supplied in the SDSS DR5 for the work presented here. For each MSG we apply 10 perturbations to the features in the training set and 10 to the blind test set, giving 100 photometric redshift values per galaxy. For LRGs and quasars, we similarly apply 32 perturbations, giving 1024 values per object. The same numbers of perturbations are used for SDSS, SDSS+GALEX, and GALEX-SDSS-only.

\begin{figure}
\figurenum{2}
\plotone{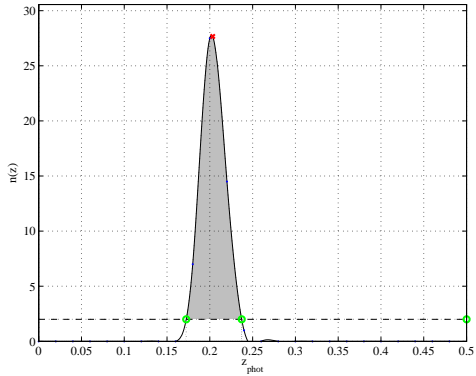}
\caption{Example PDF for an SDSS DR5 main sample galaxy. The black line is the PDF, a spline fit to the binned redshifts for each object (100 for main sample galaxies, 1024 otherwise); the red cross is the peak redshift, corresponding to half of the peak area; the blue dots are the binned individual redshifts; the horizontal dashed line is the peak threshold, and the shaded areas are the PDF peaks. \label{Fig: MSG PDF}}
\end{figure}

\begin{figure}
\figurenum{3}
\plotone{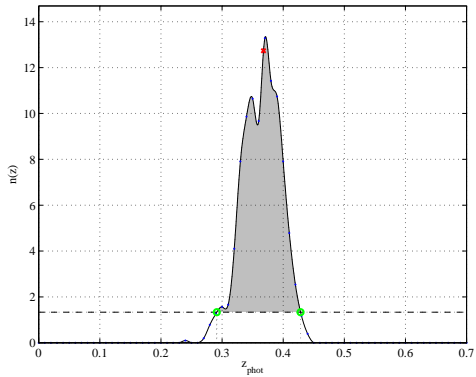}
\caption{As Figure \ref{Fig: MSG PDF} but for a luminous red galaxy. \label{Fig: LRG PDF}}
\end{figure}

\begin{figure}
\figurenum{4}
\plotone{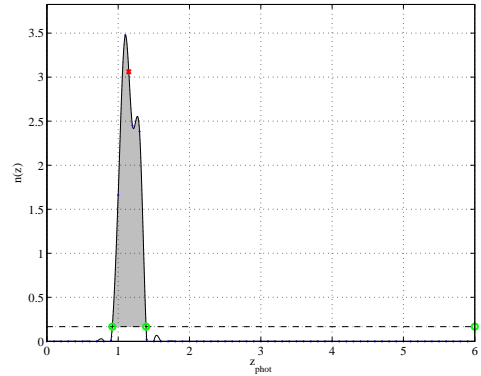}
\caption{As Figure \ref{Fig: MSG PDF} but for a quasar with one peak redshift. \label{Fig: QSO one peak PDF}}
\end{figure}

\begin{figure}
\figurenum{5}
\plotone{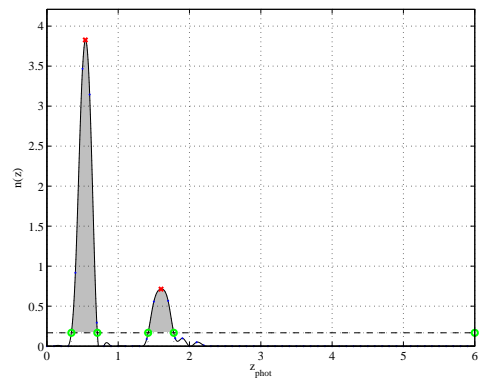}
\caption{As Figure \ref{Fig: MSG PDF} but for a quasar with multiple peak redshifts. \label{Fig: QSO multi peak PDF}}
\end{figure}

We analyze the PDFs in terms of peaks in the probability. We establish these peaks by fitting a piecewise polynomial spline to the binned redshift counts. Peak redshifts are defined as the value at which half the integrated area lies under a peak. We set a threshold under which the area does not count, so that very low peaks are not identified. The threshold level is that which the PDF would be at if it were completely flat. Thus peaks above this represent excess probability, and an object is essentially guaranteed to have at least one redshift peak. The area under a peak but also under the threshold is not included in the integral, so that the redshift values of the peaks are not pulled towards the peak centers by probability that would not otherwise count. Figures \ref{Fig: MSG PDF}-\ref{Fig: QSO multi peak PDF} show typical examples of object PDFs.

%ADM: mention removing one filter and seeing how it affects errors. NMB: Have mentioned no subset improves, but haven't e.g. quantified removing u.

\section{Results} \label{Sec: Results}

In general, we find the expected result that, for the SDSS, main galaxies and luminous red galaxies show mainly Gaussian-like PDFs, and quasars give a higher incidence of catastrophic failures. The addition of GALEX data does not greatly affect the galaxies, but substantially improves the results for quasars. We therefore present our galaxy results first, and then concentrate on the quasar results, for which we perform additional tests.

The $\zphot$ versus $\zspec$ dispersions and percentages of objects within $\dzrange$ are tabulated for the whole blind test samples in Table \ref{Table: All}, and for objects with a  single PDF peak in Table \ref{Table: One peak}. Note that the dispersions are given as $\sigma$ throughout and not $\sigma / (1+z)$.

\subsection{Galaxies} \label{Subsec: Galaxies}

\begin{figure}
\figurenum{6}
\plotone{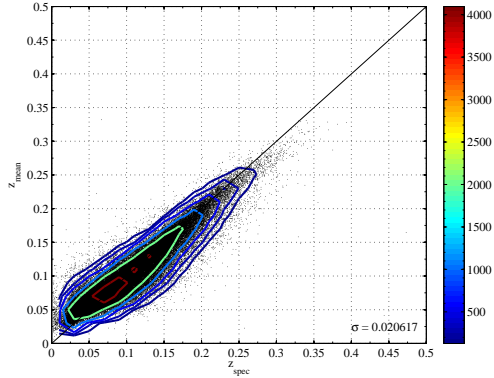}
\caption{Photometric versus spectroscopic redshift for the 82,672 SDSS DR5 main sample galaxies of the blind testing set (20\% of the sample). $\zphot$ is the mean photoz from the PDF for each object. The result from a single split (of the ten used for validation) of the data into training and blind testing data is shown. $\sigma$ is the RMS dispersion between $\zphot$ and $\zspec$. \label{Fig: zp vs zs main}}
\end{figure}

Figure \ref{Fig: zp vs zs main} shows the photometric redshift, $\zphot$, versus spectroscopic redshift, $\zspec$, for the 82,672 SDSS DR5 MSGs in the blind testing set. For each object shown, the value of $\zphot$ is the mean of the set of individual redshift values that make up its PDF. The plot shows only one instance of the training:blind split from the ten used in the holdout validation, as each split divides the sample into different subsets. Also, uniquely among the datasets, the SDSS MSGs produced a small fraction (0.2\%) of objects that could not be fit by the spline. These are excluded from the plot. We do not expect that the missing objects would have any significant impact on the results quoted here because (1) the raw PDFs were examined visually for these objects and did not appear unusual; and (2) earlier results based on direct examination of the PDF histogram included these missing objects and did not yield a significantly different value of $\sigma$.

For DR5 MSGs, we find that the overall RMS dispersion between $\zphot$ and $\zspec$, taking into account the holdout validation, is $\sigma = 0.0207 \pm 0.0001$. This is very similar to numerous previous published results, who all obtain $0.02 \lesssim \sigma \lesssim 0.025$, e.g., \citet{brunner:photoz} from Galactic fields at high latitude, and in the SDSS, \citet{tagliaferri:nnphotoz,csabai:edrphotoz,firth:annphotoz,ball:ann,collister:annz,vanzella:hdfannphotoz,wadadekar:photoz,way:photoz,abrusco:sdssphotoz,kurtz:photoz,li:photoz,oyaizu:photoz,wang:kernelphotoz,wang:novelphotoz}, and \citet{wray:photoz}. Other methods for selecting a redshift from the PDFs, for example, the mode, median, and the same values using the binned data rather than the raw redshifts, give very similar results.

The addition of GALEX data reduces the blind testing sample size to 11,969 for FUV+NUV and 20,165 for FUV-only. For NUV-only, the dispersion is the same as the optical ($\sigma = 0.0209 \pm 0.0002$), and for FUV+NUV it is slightly worse, at $\sigma = 0.0231 \pm 0.0004$. The GALEX-SDSS-only values also show very similar dispersion, indicating that GALEX photometry is making little difference to the spread. In fact, it is not surprising that the addition of the GALEX bands does little to help, for both MSG and LRG, because the dominant source of redshift information in color space, the $4000\AAi$ break, is always redwards of the GALEX bands.

%ADM: FUV+NUV maybe worse because UV detection is a function of z. NMB: add if referee asks

If one analyzes the $\zphot$ values from a single run of the NN algorithm, without perturbing the input features and taking the mean, then the dispersion is higher than that derived from the mean of the PDF, typically $\sigma \sim 0.03$. However, the galaxies become much more symmetrically distributed about the $\zphot = \zspec$ locus. We discuss possible reasons for this in \S \ref{Subsec: Galaxies Discussion}.

In the full results, if one selects galaxies with a single PDF peak, the $\sigma$ value improves slightly to to $\sigma = 0.0198 \pm 0.0001$, but the values for multiple peaks are significantly higher. For galaxies with multiple peaks, if one artificially selects the peak nearest to $\zspec$, regardless of its relative height compared to the other peaks (i.e., the `best peak', an estimate of the best possible photoz prediction), the resulting dispersions remain similar to the single-peak value for both SDSS and SDSS+GALEX.

Given that the dispersion for the whole sample is $\sigma \sim 0.02$, consistent with numerous previous results in the literature, and the generally Gaussian nature of the PDFs, and the lack of improvement from either single peaks or the artificial best peaks, we conclude that the PDFs for MSGs are approximately optimal, given the data.

\begin{figure}
\figurenum{7}
\plotone{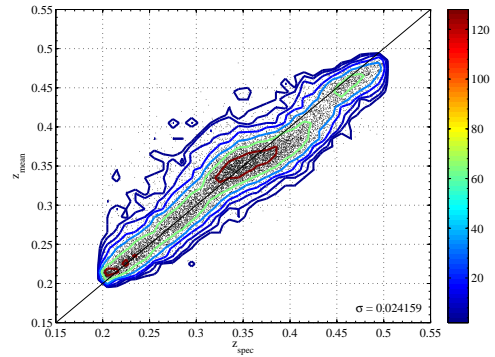}
\caption{As Figure \ref{Fig: zp vs zs main}, but for 13,254 SDSS DR5 luminous red galaxies. \label{Fig: zp vs zs LRG}}
\end{figure}

Figure \ref{Fig: zp vs zs LRG} shows the analogous results to Figure \ref{Fig: zp vs zs main} for LRGs. Again the $\sigma$ value is similar to previous work, e.g., \citet{padmanabhan:photoz,collister:megazlrg,abrusco:sdssphotoz}, and \citet{lopes:photoz}; and the single and `best' peaks again make little difference. There is a kink in the $\zphot$ versus $\zspec$ plot at $z\sim 0.35$, likely due to the movement of the $4000\AAi$ break between filters (see \S \ref{Subsec: Galaxies Discussion}).

\subsection{Quasars} \label{Subsec: Quasars}

\subsubsection{SDSS DR5} \label{Subsubsec: SDSS Quasars}

\begin{figure}
\figurenum{8}
\plotone{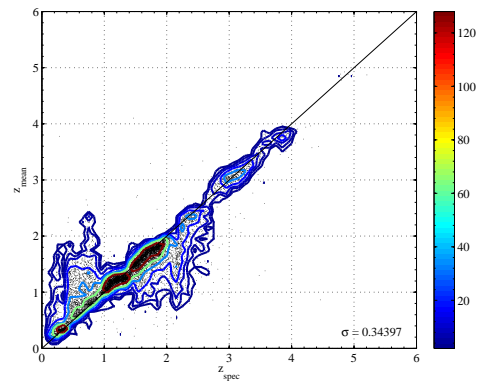}
\caption{As Figure \ref{Fig: zp vs zs main}, but for 11,149 SDSS DR5 quasars. \label{Fig: zp vs zs mean}}
\end{figure}

Figure \ref{Fig: zp vs zs mean} shows the mean PDF $\zphot$ versus $\zspec$ for 11,149 blind test quasars, in a similar manner to Figures \ref{Fig: zp vs zs main} and \ref{Fig: zp vs zs LRG}. The result is similar to that obtained by B07 using our multiple-nearest neighbor approach. We obtained a value of $\sigma = 0.35$ (there quoted as $\sigma^2 = 0.123 \pm 0.002$), compared to $\sigma = 0.343 \pm 0.005$ here.

\begin{figure}
\figurenum{9}
\plotone{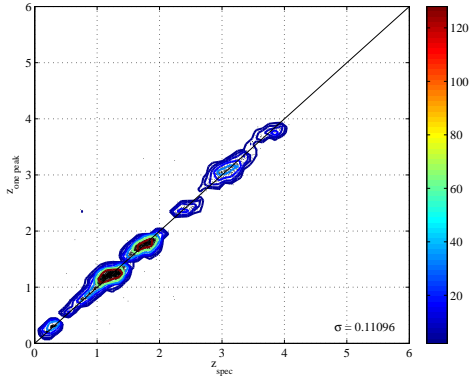}
\caption{As Figure \ref{Fig: zp vs zs main}, but for SDSS DR5 quasars with single PDF peaks. Over the ten validation runs, the number of objects with one peak from the blind test sample of 11,149 is $4339 \pm 24$. The alteration of the selection function, $n(z)$, is clear, but so is the dramatic improvement in the dispersion of the remaining objects. $99.3\%$ are within $\dz = 0.3$. \label{Fig: zp vs zs single peak}}
\end{figure}

Here, we are able to improve on this using the information available in the PDFs. Figure \ref{Fig: zp vs zs single peak} shows the same data as Figure \ref{Fig: zp vs zs mean}, but for those objects which have a single PDF peak. Although, averaged over the ten holdout validation runs, this reduces the sample size from 11,149 to $4339 \pm 24$, and alters the selection function, the improvement is dramatic. The dispersion is improved from $\sigma = 0.343 \pm 0.005$ to $\sigma = 0.117 \pm 0.001$. The percentage of quasars within $\dzrange$ is increased from $53.8 \pm 0.4\%$, $72.4 \pm 0.3\%$, and $79.8 \pm 0.3\%$ \footnote{\citet{ball:ibphotoz} found $54.9 \pm 0.7\%$, $73.3 \pm 0.6\%$, and $80.7 \pm 0.3\%$. This is consistent within the errors.} to $73.6 \pm 0.6\%$, $96.3 \pm 0.1\%$, and $99.3 \pm 0.1\%$. Just $33 \pm 4$ objects from $4339 \pm 24$ (0.7\%) remain as catastrophics. \citet{weinstein:sdssqsophotoz} obtained 83\% of quasars within $\dzdef < 0.3$ for their whole sample, but their dispersion is much higher (cf. their figure 4 and Figure \ref{Fig: zp vs zs mean}).

\begin{figure}
\figurenum{10}
\plotone{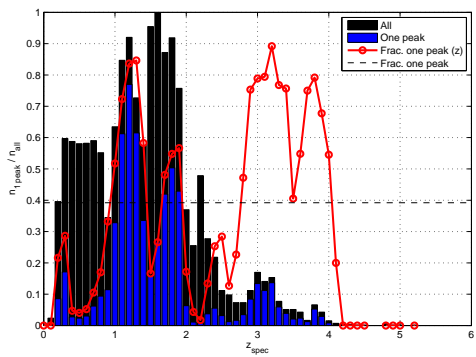}
\caption{Alteration in the selection function for the subsample of SDSS DR5 quasars with one peak compared to the full sample. The horizontal dashed line shows the overall fraction of quasars with single-peaked PDFs, to which the red line would correspond if there were no alteration. \label{Fig: One peak selection}}
\end{figure}

Figure \ref{Fig: One peak selection} shows the alteration in the selection function if we insist on only one peak in the PDF. The fraction of objects with one peak is either significantly decreased or increased from the average. Significant deficits occur at $0 \lesssim z \lesssim 1$ and $1.9 \lesssim z \lesssim 2.8$, with excess in the remaining redshift ranges. These ranges correspond to the an increased dispersion of $\sigma \sim 0.5$. We discuss possible reasons why these redshift ranges are poor in \S \ref{Subsec: Quasars Discussion}.

\subsubsection{SDSS DR5 + GALEX GR3} \label{Subsubsec: GALEX Quasars}

\begin{figure}
\figurenum{11}
\plotone{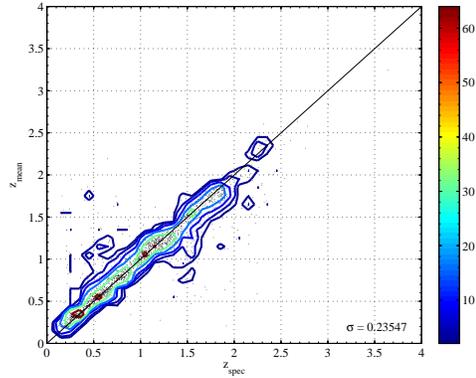}
\caption{Photometric versus spectroscopic redshift for 2066 SDSS DR5 quasars incorporating near- and far-UV photometry from matching to GALEX GR3. The improvement over SDSS alone is achieved without requiring single-peaked PDFs. \label{Fig: zp vs zs GALEX}}
\end{figure}

Figure \ref{Fig: zp vs zs GALEX} shows $\zphot$ versus $\zspec$ for the mean of the PDF in a similar manner to Figures \ref{Fig: zp vs zs main}, \ref{Fig: zp vs zs LRG}, and \ref{Fig: zp vs zs mean}. The sample size is reduced from 11,149 to 2066, but, as shown in B07, the addition of the two GALEX bands (FUV and NUV) substantially improves the results for the remaining objects. Here, the dispersion is reduced from the SDSS value of $\sigma = 0.343 \pm 0.005$ to $0.234 \pm 0.011$, and the percentage of non-catastrophics is increased from $79.8 \pm 0.3\%$ to $90.8 \pm 0.5\%$. This improvement is achieved without requiring a single peak in the PDF. When this requirement is made, the sample size is $1093 \pm 24$, the dispersion is further improved to $\sigma = 0.106 \pm 0.016$, and $99.5 \pm 0.2\%$ of the quasars are within $\dz < 0.3$.

We also find that, in addition to this FUV+NUV match, the NUV-only match produces results which are almost as good, and for a sample that is 70\% larger, at 3422 objects. The dispersion and percent non-catastrophics are $\sigma = 0.242 \pm 0.009$ and $90.8 \pm 0.4\%$. The results for the 1864 single-peaked objects are $\sigma = 0.109 \pm 0.010$ and $99.4 \pm 0.1\%$. This is a similar fraction of objects with a single peak as seen in the SDSS sample, with similar statistics.

An analysis of the GALEX-SDSS-only data confirm that the improvement, as in B07, is genuinely due to the addition of the GALEX data and not simply a consequence of requiring the objects to be detected by GALEX. The full results are given in Table \ref{Table: All}.

\section{Discussion} \label{Sec: Discussion}

\subsection{Galaxies} \label{Subsec: Galaxies Discussion}

For galaxies, the results are similar to previous work, as described in \S \ref{Subsec: Galaxies}, except that we are now presenting PDFs for each galaxy in the form of 100 (for MSG) and 1024 (for LRG) photometric redshift estimates. These PDFs can be used to improve scientific analyses of large galaxy samples.

The overall redshifts from taking the mean values of the PDFs show similar trends to previous results, with few catastrophic failures and a smoothly decreasing incidence of objects at increasing $\dzdef$. The shapes of the PDFs generally appear Gaussian, with few widely-spaced peaks. As in previous work, Figure \ref{Fig: zp vs zs main} shows that for MSGs the mean values of $\zphot$ have a slight bias toward high values at $\zspec \lesssim 0.1$. The single values of $\zphot$ from the unperturbed sample do not suffer from this and are symmetrically distributed about the $\zphot = \zspec$ locus, but the RMS dispersion is higher, at $\sigma = 0.0284 \pm 0.0002$.

%"It is likely that this behavior is due to the photometric redshifts being dominated by a single feature, in this case the 4000\AAi break.'' NMB: Prefer the parameter space explanation below

Similar behavior with respect to symmetry is seen by \citet{ball:ann} (their figure 1), who show the same biases in morphological classification, which is largely driven by the inverse concentration index. Such bias is not likely due to the lower end of the scale being zero: \citet{ball:phd} (\S 3.6.3) showed that it is still present when the targets are numerically shifted, i.e., that it is the end of the scale that matters not its numerical value. A more likely explanation is that taking the mean is analogous to using multiple neighbors in the training set. \citet{ball:ann} found that when the training set for the {\it eClass} eigenclass spectral type \citep{connolly:orthogonal,connolly:eclass,yip:eclass} is cut so that it is not dominated numerically by galaxies of the eClass corresponding to early types, the resulting classifications were spread more symmetrically about the target locus, especially in the region of early types. \citet{vanzella:hdfannphotoz} (their figure 16) show a similar example with SDSS DR1 galaxies. Likewise, we find here when using a single neighbor that the types are more symmetrically distributed, albeit with higher dispersion. Thus it is possible that using multiple neighbors is subjecting the results to the inevitable uneven distribution of objects in parameter space, in this case colors, causing the same effect as seen in \citet{ball:ann}. It appears to be a generic feature of single nearest-neighbor models (also seen for quasars) that the single neighbor produces a roughly symmetrical distribution about $\zphot = \zspec$, but using multiple neighbors (as in B07), or, in this case, taking the mean of the PDF, reduces the dispersion but introduces structure into the relation. However, in the latter case the structure is not usually large compared to the dispersion.

The LRGs, being constrained to be in the redshift range $0.2 < \zspec < 0.5$ appear to suffer less from this type of bias at low and high redshift, but instead show a kink at $\zspec \sim 0.35$. This is likely due to the $4000\AAi$ break passing between the SDSS $g$ and $r$ bands. Again, the single photoz values do not show this bias, but the dispersion is higher at $\sigma = 0.0318 \pm 0.0002$.

%More on why g to r has the bias effect if RB/ADM/referee ask

\subsection{Quasars} \label{Subsec: Quasars Discussion}

\begin{figure*}
\figurenum{12}
\plotone{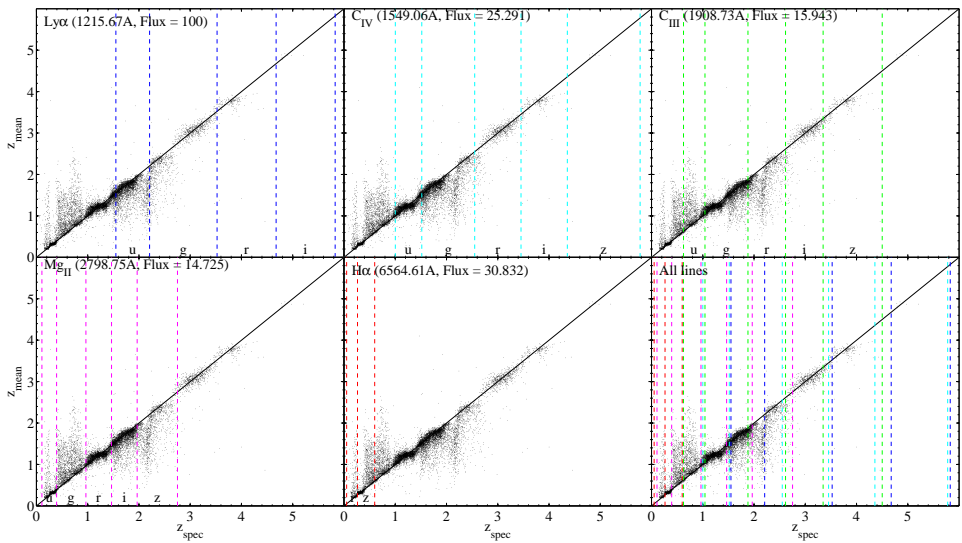}
\caption{Redshifted filters overplotted on $\zphot$ versus $\zspec$ for SDSS DR5 quasars for the five brightest emission lines. There is a clear correspondence between the redshifts at which the emission lines cross SDSS filters and the occurrence of structure in the plot. The bottom right-hand panel shows all five lines superimposed. Visually, there is no significant structure that does not correspond to one of these lines, and often several lines are close in redshift. \label{Fig: Filter}}
\end{figure*}

As described in \S \ref{Sec: Results}, there is a higher incidence of catastrophic failures among quasars compared to galaxies, which results in more PDFs with widely spaced peaks. Possible contributory factors include reddening, contamination of the quasar light by a host galaxy, degeneracy in the color-redshift relation, low equivalent-width lines, and unusual spectral slopes. But, in particular, bright spectral lines dropping between filters, or simulating other lines at a different redshift, are a major contributor. In Figure \ref{Fig: Filter}, we overplot, on the $\zphot$ versus $\zspec$ plot, the redshifts at which the five brightest emission lines of the composite quasar spectrum of \citet{vandenberk:compqsospec} cross the SDSS filter edges. The lines are, in order of flux: Ly$\alpha$, \ion{C}{4}, \ion{C}{3}, \ion{Mg}{2}, and H$\alpha$. There is a very clear correspondence between the redshifts at which the emission lines cross the filters, with particularly striking examples for \ion{Mg}{2} at $z \simeq 0.4$, H$\alpha$ at $z \simeq 0.25$, and Ly$\alpha$ at $z \simeq 2.2$. The lower right panel overplots the five lines, showing that there is no visually significant structure that does not correspond to one of these lines. The general pattern is that in between the lines, the redshifts are less spread out, and they then jump to a new value where the lines cross. It is also noticeable that most of these discontinuities not only correspond to a line crossing a filter, but to several lines doing so in a small redshift range. All this shows that objects moving between filters, and the resulting missing information or degeneracy, is a likely cause of much of the remaining error in optical quasar photometric redshifts. This information could be used to develop `optimal' filter sets for quasar photometric redshift estimation with future surveys.

%\begin{landscape}
%\end{landscape}

For multiply-peaked PDFs, when the mean $\zphot$ or $\zphot$ of the highest peak is not correct, one of the other peaks is often close to the correct redshift. It turns out that if one artificially selects these peaks, then the results are no better than those for single peaked objects (\S \ref{Subsec: Galaxies} and \S \ref{Subsubsec: SDSS Quasars}). Nevertheless, given that these objects are a smaller subset of the full sample with a different selection function, we attempted to improve the probability of the correct peak being chosen for the whole sample by applying known prior information to the derived PDFs. This was in the form of the known redshift and magnitude distributions of quasars, and their luminosity function constructed using the values of \citet{richards:dr3qsolf}. Given the known $n(z)$ of spectroscopic quasars, the photometric redshift is more likely to be in a region of high $n(z)$, thus weighting the PDF accordingly may increase the incidence of the highest peak being the correct one. Similarly, a quasar may be assigned a redshift that, given its apparent magnitude, would make it unrealistically faint or luminous. This can be downweighted by applying the magnitude distribution or luminosity function. However, we found that  the prior information did not improve the redshift statistics. This is not especially surprising because an empirical training algorithm such as the one we are using implicitly takes into account the priors by its use of the training set. Thus further weighting will not add as much information as it would, for example, in a template-based method.

Finally, we investigated the effect of altering the threshold (\S \ref{Sec: Algorithms}) above which the area under the PDF is counted as part of the redshift peaks. The main effect is to alter the relative quality of the subset of objects with single peaks. A higher threshold value will produce more objects with one peak but the sample is lower quality, and vice-versa. The threshold quoted is chosen because it corresponds to what a flat PDF would be, but it also appears approximately optimal when the percentage of objects with one peak is compared to their dispersion or $\dz$. For much higher thresholds, $\sigma$ increases relatively faster than the sample size; for much lower thresholds, the dispersion does not decrease as fast as the sample size.

From the investigations presented in this section, we conclude that, for SDSS DR5 optical data, it is unlikely that the MSG, LRG, or quasar redshifts can be significantly improved without the addition of new data. However, we have shown that, for quasars, if one is willing to discard the percentage of the objects that have more than one PDF peak, the photometric redshifts are significantly improved.

\section{Conclusions} \label{Sec: Conclusions}

We apply nearest neighbor machine learning to objects with spectra in the Sloan Digital Sky Survey Data Release 5 (SDSS DR5). We subdivide the objects into 413,361 main sample galaxies, 66,268 luminous red galaxies, and 55,743 quasars, both in the SDSS, and matched to the Galaxy Evolution Explorer All Sky Imaging Survey Third Data Release (GALEX AIS GR3). Each sample is divided into a training set consisting of 80\% of the objects, and a blind testing set of the remaining 20\%. The algorithm assigns a full probability density function (PDF) in photometric redshift to each object in the blind testing set by perturbing the input features describing the objects, in this case the colors, according to the magnitude errors. For main sample galaxies (MSGs), each PDF is formed from 100 photometric redshift values, and for luminous red galaxies (LRGs) and quasars, 1024 values.

We use the spectroscopic redshifts to test the utility of the method and find that the RMS dispersions between the photometric and spectroscopic redshifts are $\sigma = 0.0207 \pm 0.0001$, $\sigma = 0.0243 \pm 0.0002$, and $\sigma = 0.343 \pm 0.005$ for MSGs, LRGs, and quasars, respectively. The quoted errors are generated from ten-fold repeated holdout validation in the form of ten different training-to-blind-testing set splits of the data. Galaxy values are similar to previous studies, and quasar values are consistent with \citet{ball:ibphotoz}. Cross-matching to GALEX reduces the dispersion for quasars to $\sigma = 0.234 \pm 0.011$ for the 10,328 matching objects. The improvement is due to the GALEX photometry and not simply an artifact of requiring the objects to be detected by GALEX. It may be possible to improve the galaxy results by incorporating morphological information into the training set, for example the inverse concentration index, but the improvement would likely be small for these data.

For quasars, use of the PDFs enables us to construct subsamples which show dramatically improved statistics. In particular, selection of objects with a single PDF peak in the full SDSS reduces the sample size by two thirds but improves the dispersion from $\sigma = 0.343 \pm 0.005$ to $\sigma = 0.117 \pm 0.010$, with a substantial increase in the number of non-catastrophic failures (photometric minus spectroscopic redshift less than 0.3) from $79.8 \pm 0.3$\% to $99.3 \pm 0.1$\%. The equivalent statistics for the GALEX sample are $\sigma = 0.234 \pm 0.011$ to $\sigma = 0.106 \pm 0.016$, and $90.8 \pm 0.5$\% to $99.5 \pm 0.2$\%. The improved samples alter the selection function, but there is a good fraction of high percentage one-peak-regions over almost the whole redshift range.

We attempted weighting the PDFs according to known prior information on the distributions of quasars in redshift, apparent magnitude, absolute magnitude from the photometric redshift, and absolute magnitude from the luminosity function of \citet{richards:dr3qsolf}, but this did not improve the statistics. We also derived statistics for the PDF peak closest to the spectroscopic redshift regardless of its height, and found that these do not improve on the statistics for quasars with one peak. We overplot the redshifts at which bright quasar emission lines cross the SDSS filter edges and find that there is a clear correspondence with changes in the photometric redshift dispersion. This strongly suggests that a large fraction of the remaining poor redshifts are caused by lines disappearing at filter edges, or simulating other lines. We conclude that further improvement requires better data in the form of spectra for fainter objects, or a larger number of filters, both within and beyond the optical range (in particular the UV and IR).

The NN method is conceptually simple, and, once the dataset has been selected, has no adjustable parameters. This means that, unlike most machine learning algorithms, all of the information in the training data is used, and all of the computation contributes to the final result, rather than exploring parameter space and generating mostly unused results.

Future work includes the application of the methods here to more and deeper optical data, e.g., the full SDSS photometric database, the 2QZ, 2SLAQ, SDSS Southern, VVDS, DEEP2, and COSMOS surveys, and the addition of infrared data via UKIDSS and S-COSMOS. For quasars, a further useful addition would be a filter set that is customized for quasars rather than the stars and galaxies typical of optical surveys, or more and narrower filters, as in COMBO-17 \citep[e.g.,][]{wolf:combo17}, but for a wider field. The filters should overlap to minimize errors from line movement across filter edges.

%ADM: quantify expected improvement from adding IR e.g. S-COSMOS; has anyone done COMBO-17 quasar redshifts? NMB: haven't found any numbers; are only 100 quasars in COMBO-17

\acknowledgments

We thank the referee for a prompt and useful report.

The authors acknowledge support from NASA through grants NN6066H156 and NNG06GF89G, from Microsoft Research, and from the University of Illinois. The authors made extensive use of the storage and computing facilities at the National Center for Supercomputing Applications and thank the technical staff for their assistance in enabling this work.

Funding for the SDSS and SDSS-II has been provided by the Alfred P. Sloan Foundation, the Participating Institutions, the National Science Foundation, the U.S. Department of Energy, the National Aeronautics and Space Administration, the Japanese Monbukagakusho, the Max Planck Society, and the Higher Education Funding Council for England. The SDSS Web Site is http://www.sdss.org/.

The SDSS is managed by the Astrophysical Research Consortium for the Participating Institutions. The Participating Institutions are the American Museum of Natural History, Astrophysical Institute Potsdam, University of Basel, Cambridge University, Case Western Reserve University, University of Chicago, Drexel University, Fermilab, the Institute for Advanced Study, the Japan Participation Group, Johns Hopkins University, the Joint Institute for Nuclear Astrophysics, the Kavli Institute for Particle Astrophysics and Cosmology, the Korean Scientist Group, the Chinese Academy of Sciences (LAMOST), Los Alamos National Laboratory, the Max-Planck-Institute for Astronomy (MPA), the Max-Planck-Institute for Astrophysics (MPIA), New Mexico State University, Ohio State University, University of Pittsburgh, University of Portsmouth, Princeton University, the United States Naval Observatory, and the University of Washington. %Dec 19th 2005

Based on observations made with the NASA Galaxy Evolution Explorer. GALEX is operated for NASA by the California Institute of Technology under NASA contract NAS5-98034. %GALEX Cycle 2 award letter, Sep 30th 2005

Data To Knowledge (D2K) software, D2K modules, and/or D2K itineraries, used by us, were developed at the National Center for Supercomputing Applications (NCSA) at the University of Illinois at Urbana-Champaign.

This research has made use of NASA's Astrophysics Data System.

%\bibliographystyle{/Users/nball/latex/astronat/apj/apj.bst}
%\bibliography{/Users/nball/Documents/latex/refs/refs}

%Tables
\clearpage
\begin{landscape}
%\tabletypesize{\scriptsize}
\begin{deluxetable*}{ccccccccc}
\tabletypesize{\tiny}
%\rotate
\tablenum{1}
\tablewidth{8.5in}
\tablecaption{Photometric redshift PDF statistics. For quasars, the improvements resulting from cross-matching to the GALEX UV bands are clear. For SDSS main sample galaxies (MSGs) and luminous red Galaxies (LRGs), the cross-match produces little improvement, as expected. Quoted errors are the standard deviation from ten-fold repeated holdout validation. The $\dz$ ($\dzdef$) thresholds are 0.01, 0.02, and 0.03 for MSGs and LRGs, and 0.1, 0.2, and 0.3 for quasars. \label{Table: All}}
\tablehead{\colhead{Dataset} &\colhead{Objects} &\colhead{Subset} &\colhead{$\mathrm{Sample Size_{all}}$} &\colhead{$\mathrm{RMS_{all}}$} &\colhead{\parbox[b]{0.75in}{\centering $\dz < 0.01,0.1$\\(\%)}} &\colhead{\parbox[b]{0.75in}{\centering $\dz < 0.02,0.2$\\(\%)}} &\colhead{\parbox[b]{0.75in}{\centering $\dz < 0.03,0.3$\\(\%)}}} %&\colhead{$\dz < 0.05/0.5$ (\%)}}
\startdata
 SDSS            &MSG &-        &82,672 &$0.0207 \pm 0.0001$ &$44.8 \pm 0.2$ &$72.9 \pm 0.2$ &$87.1 \pm 0.1$\\ %&$97.2 \pm 0.1$\\
 SDSS            &LRG &-        &13,254 &$0.0243 \pm 0.0002$ &$40.1 \pm 0.3$ &$68.2 \pm 0.4$ &$84.0 \pm 0.4$\\ %&$95.7 \pm 0.1$\\
 SDSS            &QSO &-        &11,149 &$0.343  \pm 0.005$  &$53.8 \pm 0.4$ &$72.4 \pm 0.3$ &$79.8 \pm 0.3$\\ %&$87.6 \pm 0.2$\\
 SDSS+GALEX      &MSG &FUV+NUV  &11,969 &$0.0231 \pm 0.0004$ &$37.0 \pm 1.0$ &$65.6 \pm 0.9$ &$83.5 \pm 0.5$\\ %&$96.4 \pm 0.2$\\
 SDSS+GALEX      &MSG &NUV only &20,165 &$0.0209 \pm 0.0002$ &$42.4 \pm 0.2$ &$71.5 \pm 0.2$ &$86.9 \pm 0.2$\\ %&$97.4 \pm 0.1$\\
 SDSS+GALEX      &LRG &FUV+NUV  &51     &$0.0304 \pm 0.0035$ &$31.4 \pm 4.0$ &$57.6 \pm 5.4$ &$73.1 \pm 4.5$\\ %&$88.2 \pm 3.1$\\
 SDSS+GALEX      &LRG &NUV only &463    &$0.0260 \pm 0.0018$ &$38.1 \pm 1.8$ &$66.6 \pm 1.3$ &$81.4 \pm 1.4$\\ %&$94.3 \pm 0.9$\\
 SDSS+GALEX      &QSO &FUV+NUV  &2066   &$0.234  \pm 0.011$  &$71.8 \pm 0.6$ &$86.4 \pm 0.7$ &$90.8 \pm 0.5$\\ %&$94.8 \pm 0.4$\\
 SDSS+GALEX      &QSO &NUV only &3422   &$0.242  \pm 0.009$  &$68.2 \pm 0.7$ &$85.4 \pm 0.6$ &$90.8 \pm 0.4$\\ %&$94.7 \pm 0.4$\\
 GALEX-SDSS-only &MSG &FUV+NUV  &11,969 &$0.0230 \pm 0.0004$ &$38.4 \pm 0.5$ &$67.0 \pm 0.5$ &$83.4 \pm 0.5$\\ %&$96.4 \pm 0.2$\\
 GALEX-SDSS-only &MSG &NUV only &20,165 &$0.0224 \pm 0.0003$ &$39.6 \pm 0.2$ &$68.2 \pm 0.3$ &$84.2 \pm 0.4$\\ %&$96.5 \pm 0.2$\\
 GALEX-SDSS-only &LRG &FUV+NUV  &51     &$0.0292 \pm 0.0030$ &$29.6 \pm 5.4$ &$58.0 \pm 5.6$ &$74.3 \pm 4.2$\\ %&$90.6 \pm 2.2$\\
 GALEX-SDSS-only &LRG &NUV only &463    &$0.0257 \pm 0.0018$ &$38.9 \pm 2.4$ &$66.7 \pm 1.9$ &$82.0 \pm 1.2$\\ %&$94.7 \pm 0.9$\\
 GALEX-SDSS-only &QSO &FUV+NUV  &2066   &$0.314  \pm 0.013$  &$64.2 \pm 0.9$ &$79.9 \pm 0.5$ &$85.5 \pm 0.5$\\ %&$91.0 \pm 0.4$\\
 GALEX-SDSS-only &QSO &NUV only &3422   &$0.336  \pm 0.010$  &$57.1 \pm 0.8$ &$75.1 \pm 0.8$ &$81.9 \pm 0.7$\\ %&$88.9 \pm 0.7$\\
\enddata
\end{deluxetable*}

%\tabletypesize{\scriptsize}
\begin{deluxetable*}{ccccccccc}
%\rotate
\tablenum{2}
\tablewidth{8.5in}
\tablecaption{As Table \ref{Table: All}, but for the subsamples of objects with a single peak in the redshift probability density function. The improvement is particularly dramatic for the SDSS quasar sample. Unlike the full samples, the sample size now contains a confidence interval due to the holdout validation, but the variation is small, of order 1\%. \label{Table: One peak}}
\tablehead{\colhead{Dataset} &\colhead{Objects} &\colhead{Subset} &\colhead{$\mathrm{Sample Size_{1 peak}}$} &\colhead{$\mathrm{RMS_{1 peak}}$} &\colhead{\parbox[b]{0.75in}{\centering $\dz < 0.01,0.1$\\(\%)}} &\colhead{\parbox[b]{0.75in}{\centering $\dz < 0.02,0.2$\\(\%)}} &\colhead{\parbox[b]{0.75in}{\centering $\dz < 0.03,0.3$\\(\%)}}} %&\colhead{$\dz < 0.05/0.5$ (\%)}}
\tablecolumns{8}
\startdata
 SDSS            &MSG &-        &$71,236 \pm 145$ &$0.0198 \pm 0.0009$ &$45.9 \pm 0.2$ &$74.2 \pm 0.2$ &$88.1 \pm 0.1$\\ %&$97.7 \pm 0.1$\\
 SDSS            &LRG &-        &$11,231 \pm 87$  &$0.0223 \pm 0.0002$ &$40.4 \pm 0.4$ &$69.4 \pm 0.4$ &$85.4 \pm 0.4$\\ %&$96.7 \pm 0.2$\\
 SDSS            &QSO &-        &$4339   \pm 24$  &$0.117  \pm 0.001$  &$73.6 \pm 0.6$ &$96.3 \pm 0.1$ &$99.3 \pm 0.1$\\ %&$99.7 \pm 0.1$\\
 SDSS+GALEX      &MSG &FUV+NUV  &$7276   \pm 164$ &$0.0214 \pm 0.0004$ &$38.4 \pm 1.4$ &$67.7 \pm 1.2$ &$85.4 \pm 0.6$\\ %&$97.3 \pm 0.2$\\
 SDSS+GALEX      &MSG &NUV only &$12,077 \pm 106$ &$0.0191 \pm 0.0002$ &$45.4 \pm 0.2$ &$74.8 \pm 0.3$ &$89.3 \pm 0.2$\\ %&$98.2 \pm 0.2$\\
 SDSS+GALEX      &LRG &FUV+NUV  &$6.5    \pm 2.3$ &$0.0204 \pm 0.0170$ &$41.5 \pm 19$  &$82.5 \pm 20$  &$93.5 \pm 11$ \\ %&$98.3 \pm 5.3$\\
 SDSS+GALEX      &LRG &NUV only &$187.3  \pm 8.6$ &$0.0198 \pm 0.0015$ &$43.0 \pm 3.3$ &$75.2 \pm 2.4$ &$88.7 \pm 2.3$\\ %&$97.8 \pm 1.1$\\
 SDSS+GALEX      &QSO &FUV+NUV  &$1093   \pm 24$  &$0.106  \pm 0.016$  &$83.8 \pm 0.7$ &$97.9 \pm 0.3$ &$99.5 \pm 0.2$\\ %&$99.6 \pm 0.1$\\
 SDSS+GALEX      &QSO &NUV only &$1864   \pm 32$  &$0.109  \pm 0.010$  &$80.0 \pm 0.8$ &$97.2 \pm 0.3$ &$99.4 \pm 0.1$\\ %&$99.7 \pm 0.1$\\
 GALEX-SDSS-only &MSG &FUV+NUV  &$6284   \pm 197$ &$0.0211 \pm 0.0003$ &$40.5 \pm 0.5$ &$69.8 \pm 0.6$ &$85.6 \pm 0.5$\\ %&$97.3 \pm 0.2$\\
 GALEX-SDSS-only &MSG &NUV only &$11,892 \pm 106$ &$0.0205 \pm 0.0002$ &$42.7 \pm 0.4$ &$71.7 \pm 0.4$ &$86.7 \pm 0.4$\\ %&$97.4 \pm 0.1$\\
 GALEX-SDSS-only &LRG &FUV+NUV  &$4.6    \pm 2.4$ &$0.0169 \pm 0.0075$ &$41.9 \pm 24$  &$81.6 \pm 18$  &$89.7 \pm 15$ \\ %&$100  \pm 0$\\
 GALEX-SDSS-only &LRG &NUV only &$196.4  \pm 8.8$ &$0.0206 \pm 0.0017$ &$43.6 \pm 4.7$ &$72.4 \pm 4.0$ &$87.3 \pm 2.7$\\ %&$97.3 \pm 0.8$\\
 GALEX-SDSS-only &QSO &FUV+NUV  &$933    \pm 20$  &$0.143  \pm 0.031$  &$81.5 \pm 1.2$ &$97.2 \pm 0.7$ &$98.8 \pm 0.4$\\ %&$99.1 \pm 0.4$\\
 GALEX-SDSS-only &QSO &NUV only &$1542   \pm 18$  &$0.123  \pm 0.015$  &$77.1 \pm 0.7$ &$96.7 \pm 0.4$ &$99.1 \pm 0.2$\\ %&$99.5 \pm 0.2$\\
\enddata
\end{deluxetable*}
\clearpage
\end{landscape}

\end{document}